\RequirePackage{fix-cm}
\documentclass[twocolumn,natbib]{svjour3}          
\smartqed  
\usepackage{graphicx}
\usepackage{multirow}
\usepackage{subfigure}
\usepackage{amsmath}
\usepackage{amssymb}
\usepackage{mathptmx}
\usepackage{amsfonts}
\usepackage{times}
\usepackage{color}
\usepackage{fancyhdr}
\usepackage{gensymb}
\usepackage{lineno}
\usepackage{booktabs}
\usepackage{eurosym}

\usepackage{hyperref}
\usepackage{color}

\usepackage{colortbl}
\usepackage{bm}

\usepackage{soul}

\usepackage{algorithm}
\usepackage{algorithmic}
\begin{document}

\title{A Fuzzy-PSO System for Indoor Localization based on Visible Light Communications}


\author{Giovanni Pau \and
	Mario Collotta \and
	Vincenzo Maniscalco \and
	Kim-Kwang Raymond Choo
}


\institute{G.Pau, M.Collotta, V.Maniscalco \at
	Faculty of Engineering and Architecture, Kore University of Enna, 94100 - Enna, Italy\\
	\email{\{giovanni.pau;mario.collotta;vincenzo.maniscalco\}\\@unikore.it}    
	\and
	K.K.R.Choo\at
	Department of Information Systems and Cyber Security and Department of Electrical and Computer Engineering, University of Texas at San Antonio, San Antonio, TX 78249-0631, USA\\
	\email{raymond.choo@fulbrightmail.org}
}

\date{Received: date / Accepted: date}

\maketitle

\begin{abstract}
Indoor positioning systems using Visible Light Communication (VLC) have potential applications in smart buildings, for instance, in developing economical, easy-to-use, widely accessible positioning system based on Light Emitting Diodes (LEDs). Thus using VLCs, we introduce a new fuzzy-based system for indoor localization in this paper. The system processes data from transmitters (i.e., anchor nodes) and delivers the calculated position of a receiver. A Particle Swarm Optimization (PSO) technique is then employed to obtain the optimal configuration of the proposed Fuzzy Logic Controllers (FLCs). Specifically, the proposed PSO technique optimizes the membership functions of the FLCs by adjusting their range to achieve the best results regarding the localization reliability. We demonstrate the utility of the proposed approach using experiments.
\keywords{Visible light communications \and indoor localization \and received signal strength indication \and fuzzy logic controller \and particle swarm optimization}
\end{abstract}

\section{Introduction}
\label{sec:introduction} 


Light emitting diodes (LEDs), known for their illumination efficiency, eco-friendliness, and durability (lifetime) \citep{led-survey}, are semiconductors that can be simply modulated and used in communication systems \citep{Pau201744}. Visible light communication (VLC) using white LEDs is also increasingly popular since they can operate as lighting and communications systems simultaneously \citep{vlc-survey,biagi-extra,mangold-extra,zhang-extra}. Such communication is also carried out in the license-free spectrum and produces no electromagnetic interference. Besides, VLC can be adopted in particularly sensitive areas \citep{lampe-extra}, such as airplanes and hospitals. Not surprisingly, LED-based VLCs have been proposed for sensing networks \citep{vlc-sensing}, illumination \citep{vlc-illumination}, intelligent transportation systems \citep{vlc-its}, broadcasting \citep{vlc-broadcasting}, and many other applications.


A trend in recent years is to ensure indoor location systems including self-sufficient robot management \citep{robot-localization,Prorok20113241}, position identification \citep{position-revognition}, and location-based services \citep{location-services,jisis17-7-3-02,jisis16-6-4-04}. For instance, there are several localization and positioning systems proposed in the literature, such as those based on GPS \citep{indoor-gps}, RFID \citep{indoor-rfid}, infrared \citep{indoor-infrared}, ultrasound \citep{indoor-ultrasound}, WLAN \citep{indoor-wlan}, Bluetooth \citep{indoor-bluetooth} and other approaches \citep{indoor-survey}. However, GPS may not be fit-for-purpose in indoor situations due to multipath fading (e.g., caused by objects and surfaces) and power attenuation. Indoor positioning systems based on RFID, ultrasound, WLAN, infrared, and Bluetooth also have several constraints, such as electromagnetic interference, requiring the installation of new infrastructures \citep{jowua17-8-3-03}, low certainty, relatively slow responses, and low security. Multipath propagation issues also affect these systems; thus, it is particularly challenging to ascertain the direction or the distance of the transmitter from the obtained signal. Hence, the development of innovative strategies based on existing networks is of high importance.

One particular solution is to design indoor positioning systems based on VLC \citep{vlc-survey,Yi2015436}, using the light released by LEDs \citep{xu-extra}. LED-based positioning approaches are usually economical, easy-to-use, and can be integrated into indoor localization systems. Existing VLC-based indoor positioning approaches are briefly summarized as follow:
\begin{itemize}
    \item \textit{Scene Analysis and Proximity} (fingerprinting): Fingerprinting can be applied to the VLC, i.e., the gathering of necessary information, followed by the second measurement before a real-time comparison. This technique is simple and does not require complicated processing. However, it requires a large amount of relevant information to be collected. If such information is not available, then it would lead to inaccurate estimates. This approach has been applied in visible-light beacons indoor positioning \cite{vlc-loc-fingerprint}, where a correlation-based technique is employed to decompose light signals and to obtain fingerprints. Subsequently, the authors used a localization framework to improve the precision. A comparison is carried out with other localization systems and the findings suggested that the authors' proposed (fingerprinting-based) solution does not always allow achieve the best performance.
    \item \textit{Time Difference of Arrival} (TDOA): Each LED uses a specific frequency, and by applying appropriate band-pass filters at the receiver, it is possible to detect each of them. The TDOA method is based on the same principle of the Time of Arrival (TOA) approach \citep{Wang20133302}. In the latter, the time required for a signal to arrive from a transmitter to a receiver is measured, and its distance is then calculated. Unlike TOA (that computes the propagation delay between the receiver and each transmitter), TDOA exploits the difference in propagation time between them to estimate the distance between the transmitters (whose coordinates are already known). In this case, it is necessary that only the transmitters have to be synchronized and not the receivers. A TDOA approach was proposed in \cite{vlc-loc-tdoa1} to estimate the target position by using LED ceiling lamps. The authors explained that their system can potentially be used for future indoor positioning in environments with ceiling composed of LEDs light. In a later work, \cite{vlc-loc-tdoa2} improved the approach presented in \cite{vlc-loc-tdoa1} incorporating measurement uncertainty generated by Additive White Gaussian Noise to achieve better accuracies. While TDOA method may be an appropriate solution in some contexts, it is not an optimal choice for economic LED positioning. Other approaches based on Phase Difference of Arrival (PDOA) \citep{loc-pdoa} can also be used for VLC positioning, but they also suffer from the same limitations.
	\item \textit{Angle of Arrival } (AOA): AOA is defined as the angle between the propagation path of a wave (with its incidence) and a direction of reference, which is identified as orientation. This approach has been applied in \cite{vlc-loc-aoa,little-extra}, where the authors in both works introduced new efficient and low-complexity solutions, by using VLC, for the localization of devices in indoor environments. Both findings suggested that a localization with a precision of the order of a meter could be achieved.
    \item \textit{Image positioning}: These techniques usually employ image sensors to capture images of LEDs \citep{Huynh2016}. Then, the position of the image sensor is estimated considering the correlation between the 3D coordinates of LEDs (that are known) and the 2D coordinates of LEDs in the obtained image(s) \citep{vlc-loc-image1,vlc-loc-image2}.
    \item \textit{Received Signal Strength} (RSS): The distance is estimated from the information on the pulses received from the various transmitters \citep{Jung201363}. A method based on RSS measures was proposed by \cite{vlc-loc-rss-biagi}. In this method, each LED has its carrier to reduce interference between the LEDs, while the receiver determines the distance by measuring the RSS of the LED light and, finally, estimates the position. In \cite{vlc-loc-rss1}, the authors also presented an indoor positioning method that employs a single LED array and many tilted optical receivers. Three-dimensional positioning is achieved by managing the RSS. Another indoor localization system, by utilizing VLC, for mobile robots was proposed by \cite{vlc-loc-rss2}. In this case, the positioning is obtained by employing a multi-frequency method with the RSS to estimate the distance between a robot and each LED. Findings appear to be promising. Thus, RSS-based approaches can be a viable support for indoor positioning and localization based on VLC.
	
\end{itemize}

\begin{figure*}
	\centering
	\includegraphics[width=5in]{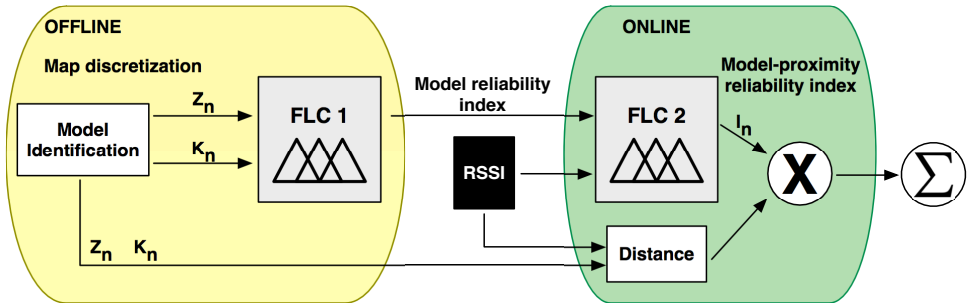}
	\caption{Architecture of the suggested system.}
	\label{fig:system-model}
\end{figure*}


In this paper, an innovative fuzzy-based localization system by using VLC is presented. Specifically, the proposed approach extends the concept of trilateration without the need to solve several equations required for determining the location of a receiver. A fuzzy-based solution is chosen, since it has been demonstrated in the literature to be a viable approach for indoor localization \citep{vlc-loc-image2,Jung201363,vlc-loc-rss-biagi}. Furthermore, the application of Fuzzy Logic Controllers (FLCs) facilitates the development of control procedures with multi-criteria. Fuzzy logic is also capable of performing real-time choices, while traditional control systems often rely on an exact description of the controlled environment that is not usually available. Considering that fuzzy logic methods can efficiently manage the linguistic rules, they can be attractive for a diverse range of applications such as indoor localization. At the time of this research, there is no solution in the literature where a VLC-based indoor localization system is supported using fuzzy logic controllers. This is the focus of this paper.


In the proposed VLC-based indoor localization system, the environment map is discretized in a reference grid; subsequently, a fuzzy-based approach is applied for anchor weighting and, finally, these weights are summed. We adopt the Particle Swarm Optimization (PSO) to achieve the best parameters and values of the fuzzy-based system. In particular, PSO is used to optimize the membership functions of the Fuzzy Logic Controller, by adjusting their range. PSO, an evolutionary computation method \citep{pso}, is recognized as a valid heuristic technique for optimization problems in multidimensional and continuous research spaces. It has also been shown that the PSO technique can be used to achieve high-quality solutions while minimizing the computational load \citep{pso-fuzzy1,pso-fuzzy2,pso-fuzzy3,pso-fuzzy4,pso-fuzzy-pau}, unlike stochastic methods such as the genetic algorithms. Although PSO has been used in FLCs optimization, we are not aware of any existing work using PSO in a VLC-based indoor localization system. 

In the next section, we will introduce the proposed system. Section \ref{sec:pso} presents the proposed PSO algorithm and how it can be used to optimize the FLC. Section \ref{sec:performance} presents the evaluation of the proposed system's performance in a testbed scenario, and Section \ref{sec:conclusions} concludes the paper.

\section{The Proposed Solution}
\label{sec:system-model}

\begin{figure}
	\centering
	\includegraphics[width=2in]{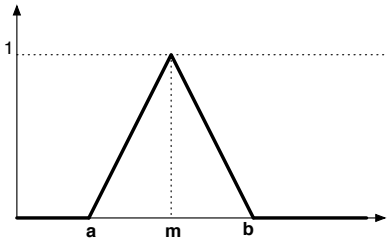}
	\caption{General representation of a triangular membership function.}
	\label{fig:triangular-gen}
\end{figure}

The proposed system is defined by the architecture represented in Figure \ref{fig:system-model}, and consists of two phases, namely: offline training and online localization, where a fuzzy inference system is employed. In the proposed system, triangular membership functions are chosen for the parameters in the fuzzy inference system. Let $x$ be the  general variable. Every membership function can be described by a general triangular-shaped mathematical description as follow:

\begin{equation} \label{eq:triagular}
\mu_A(x)=
\begin{cases} 
0 & \mbox{if }x \leq a \\ 
\dfrac{x-a}{m-a} & \mbox{if }a < x \leq m \\ 
\dfrac{b-x}{b-m} & \mbox{if }m < x < b \\ 
0 & \mbox{if }x \geq b \\ 
\end{cases}
\end{equation}
where $a$ is a lower limit, $b$ is upper limit and $m$ a value, with $a < m < b$ (Figure \ref{fig:triangular-gen}). The operation of both stages, depicted in Figure \ref{fig:system-model}, is described in the following subsections.

\subsection{Offline Stage}
\label{subsec:offline}

The main aim of the offline stage is to determine the parameters of the RSSI-distance equation, of which a simplified version, taking into account irradiance angles of the light sources and the incident angles, is the following \citep{rssi-eq,rssi-refref}:

\begin{equation} \label{eq:rssi}
RSSI = Z\cdot\log_{10}(w)+K
\end{equation}
where the $RSSI$ is estimated in power ratio while $w$ is the distance (in meters) between the receiver node and the beacon. Obviously, as in all RSSI-based approaches, even the one proposed in this paper assumes that the communication is in LOS (Line of Sight). It is required to obtain the values of $Z$ and $K$ parameters. To this end, the least squares technique is taken into account in this paper. The map of the environment is arranged into square cells with a side equal to $S$. A specific value is assigned to every single cell, and its initialization value is $0$. In the proposed solution, for each cell ($i$,$j$), the range $w_n$ ($i$,$ j$) between the center of the cell and the anchor $n$, whose positions are known, is estimated.

\begin{figure}
	\centering
	\subfigure[]
	{\includegraphics[width=1.55in]{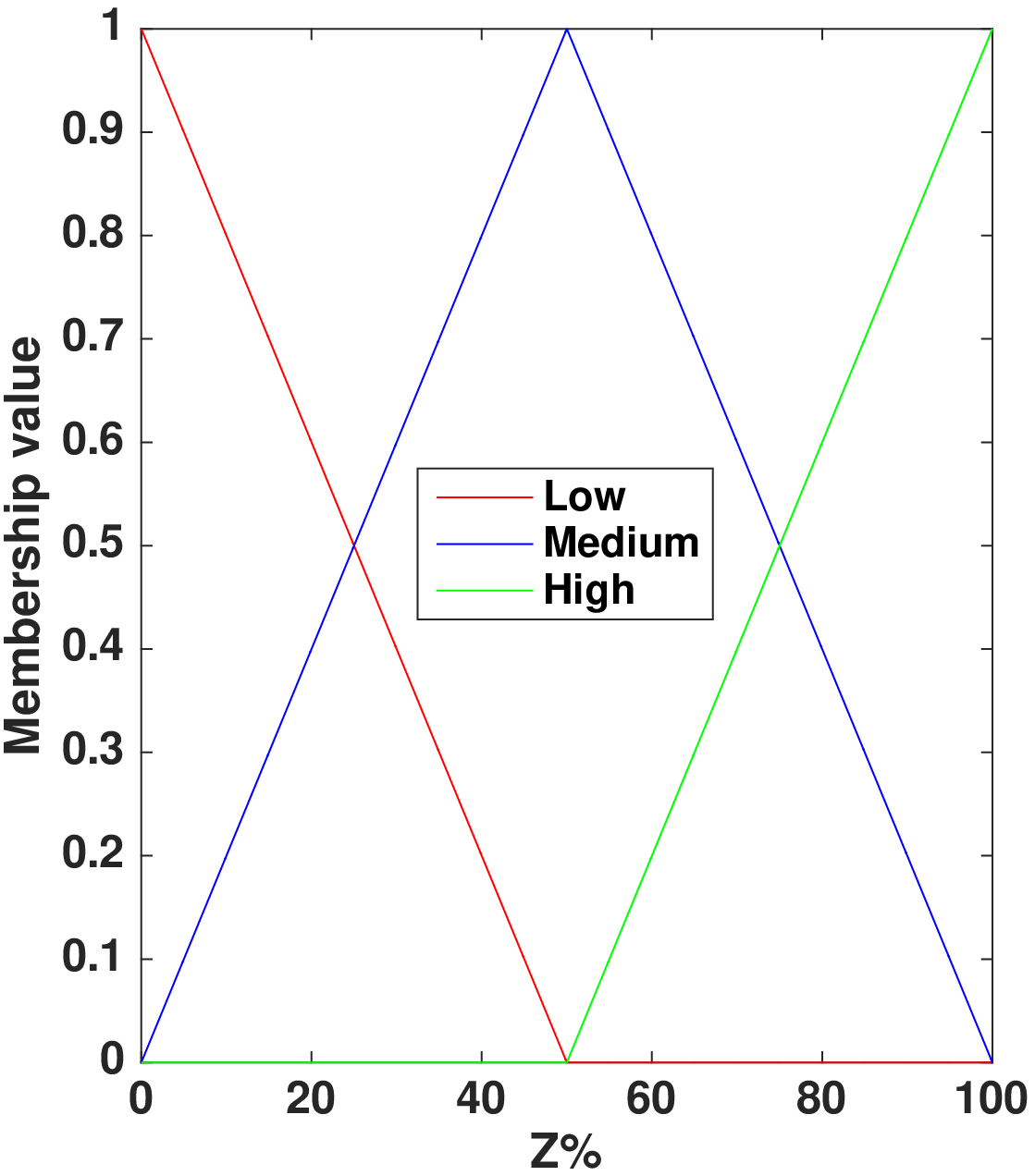}}
	\subfigure[]
	{\includegraphics[width=1.55in]{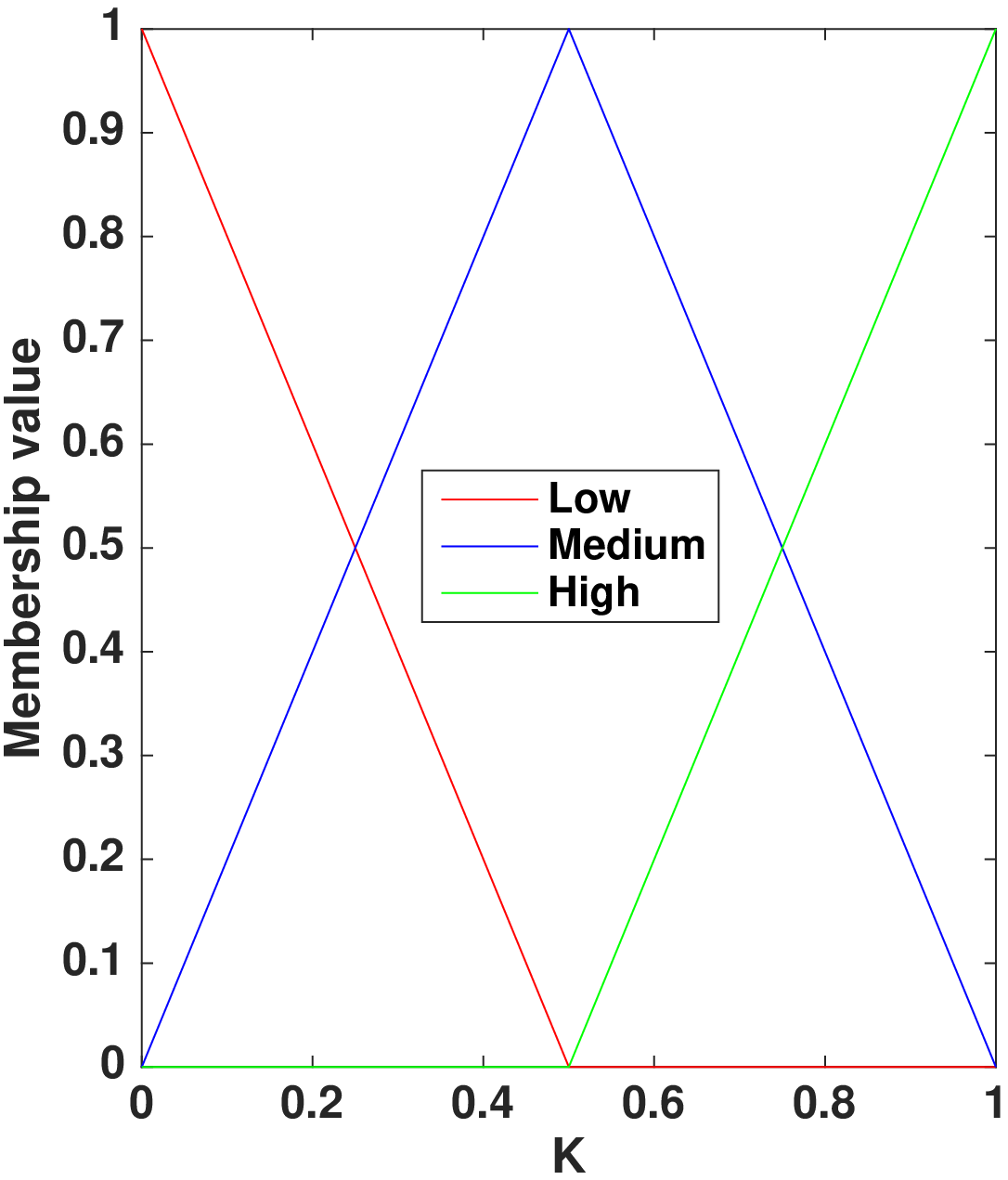}}
	\caption{Membership functions of input parameters for the FLC 1: a) $Z$; b) $K$.}
	\label{fig:mf-fuzzy1}
\end{figure}

Besides, the goal of the offline stage is to ascertain a (pre-)weighting parameter, closely related to modeling and estimating of accuracy, faulty localization, and reliability of each anchor employed in the map of the environment. In fact, this cumulative reliability index is obtained as the output of a Fuzzy Logic Controller (FLC). On the contrary, the input parameters are $Z_n$ and $K_n$, which are organized in 3 fuzzy sets with triangular membership functions. They represent the various levels of correlation between each parameter and its theoretical value. It is necessary to note that this typical value is determined by measuring the median of each earned value of $Z_n$ and $K_n$. The membership functions of $Z_n$ and $K_n$ are pictured in Figure \ref{fig:mf-fuzzy1}, where the membership is outlined by normalized values [$0\div 1$]. Furthermore, considering the equation \ref{eq:triagular}, the different values of the variables are shown in Table \ref{tab:fuzzy1}. The output value of the FLC 1, reported in Table \ref{tab:output-fuzzy1}, is determined by 9 fuzzy rules and represents the model reliability index, fluctuating from $0$ to $1$. For instance, if $Z\%$ is \textit{Medium} and $K$ is \textit{High} then the output value is $0.75$.

\begin{table}[]
	\centering
	\caption{FLC 1: values of variables used in definition of triangular membership functions.}
	\label{tab:fuzzy1}
	\begin{tabular}{|c|c|c|c|c|}
		\hline
		\textbf{Input Variable} & \textbf{Linguistic term} & \textit{\textbf{a}} & \textit{\textbf{m}} & \textit{\textbf{b}} \\ \hline
		\multirow{3}{*}{Z\%}      & Low                     & 0                   & 0                   & 50                  \\ \cline{2-5} 
		& Medium                   & 0                   & 50                  & 100                 \\ \cline{2-5} 
		& High                      & 50                  & 100                 & 100                 \\ \hline
		\multirow{3}{*}{K}      & Low                     & 0                   & 0                   & 0.5                 \\ \cline{2-5} 
		& Medium                   & 0                   & 0.5                 & 1                   \\ \cline{2-5} 
		& High                      & 0.5                 & 1                   & 1                   \\ \hline
	\end{tabular}
\end{table}

\begin{table}[]
	\centering
	\caption{FLC 1: inference rules of the model reliability index}
	\label{tab:output-fuzzy1}
	\begin{tabular}{|c|c|c|c|c|}
		\hline
		\multicolumn{2}{|c|}{\multirow{2}{*}{}}         & \multicolumn{3}{c|}{\textbf{K}}                \\ \cline{3-5} 
		\multicolumn{2}{|c|}{}                          & \textit{Low} & \textit{Medium} & \textit{High} \\ \hline
		\multirow{3}{*}{\textbf{Z\%}} & \textit{Low}    & 0.03        & 0.06            & 0.15           \\ \cline{2-5} 
		& \textit{Medium} & 0.25          & 0.45             & 0.75           \\ \cline{2-5} 
		& \textit{High}   & 0.65          & 0.85             & 1             \\ \hline
	\end{tabular}
\end{table}

\subsection{Online Stage}
\label{subsec:online}

In the proposed system, when it is necessary to locate an undiscovered node placed at the center of a generic cell in the environment map, the value related to each cell denotes an evaluation of the error that may affect the localization mechanism. The main aim is to obtain an aggregate value for each cell and, in the end, select the cell with the smallest value. To this end, the overall map of errors is realized by subsequent steps, calculated for each anchor node and for every cell:

\begin{figure}
	\centering
	\includegraphics[width=3.8in]{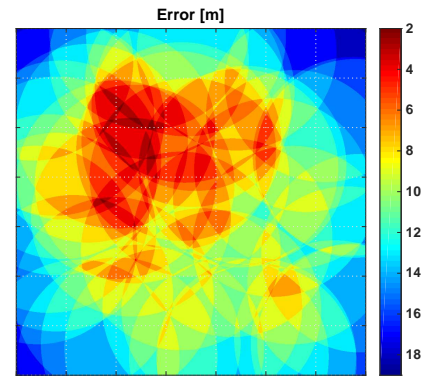}
	\caption{An example of error map generated from a generic anchor node. The side $S$ of the square cells is 1 m.}
	\label{fig:errormap}
\end{figure}

\begin{figure}
	\centering
	\subfigure[]
	{\includegraphics[width=1.55in]{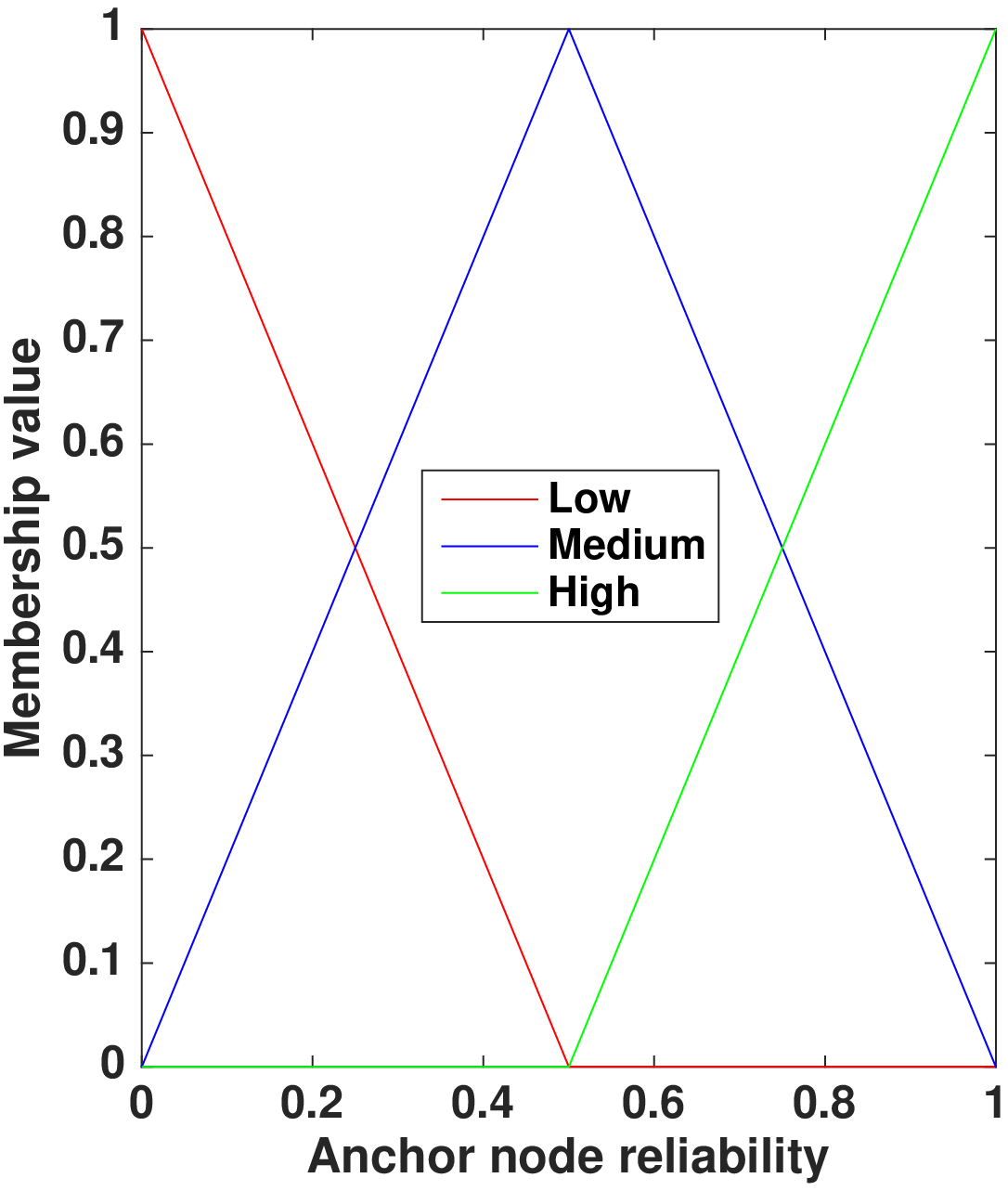}}
	\subfigure[]
	{\includegraphics[width=1.55in]{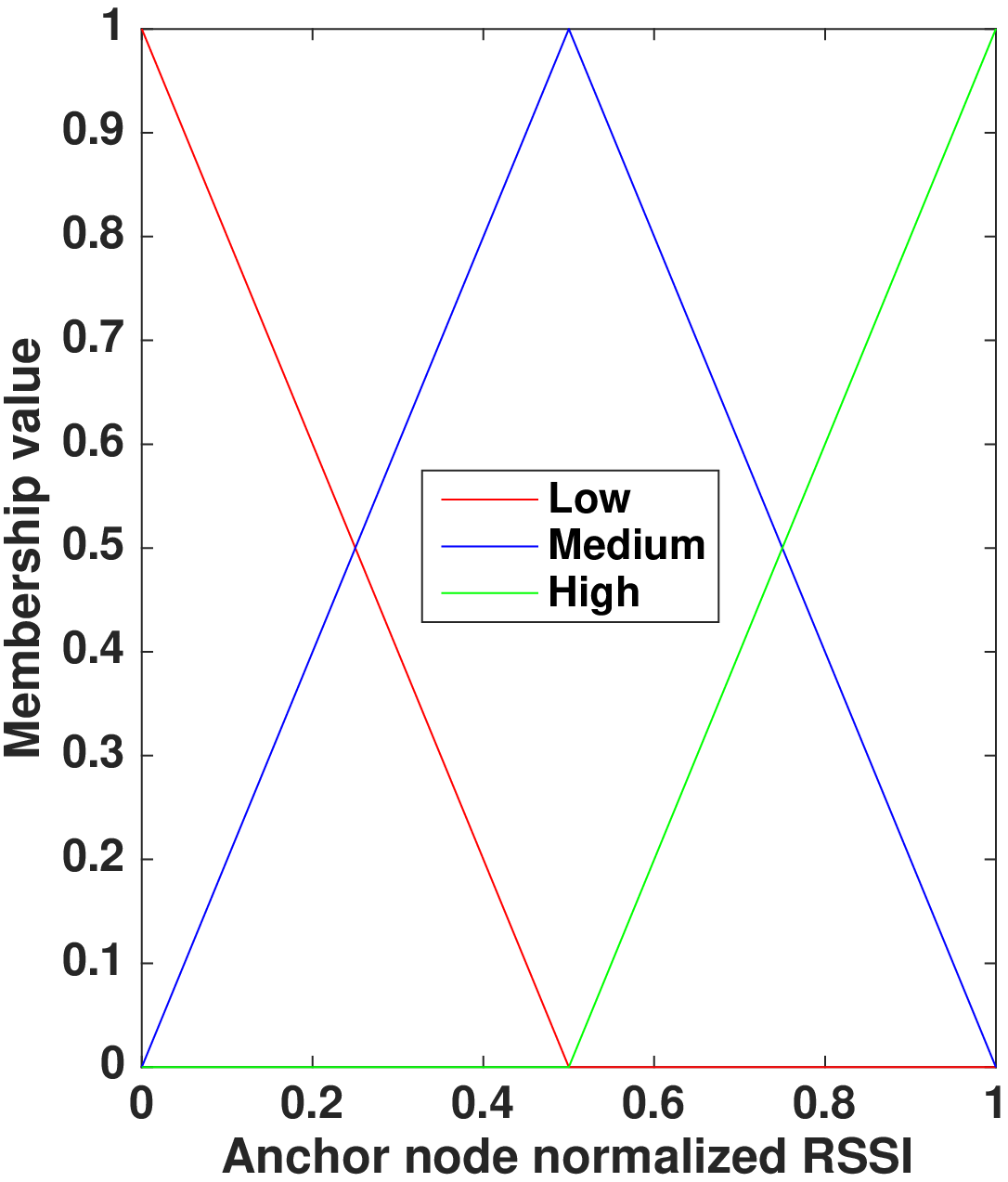}}
	\caption{Membership functions of input parameters for the FLC 2: a) Anchor node reliability; b) Anchor node normalized RSSI.}
	\label{fig:mf-fuzzy2}
\end{figure}

\begin{figure}
	\centering
	\includegraphics[width=3.5in]{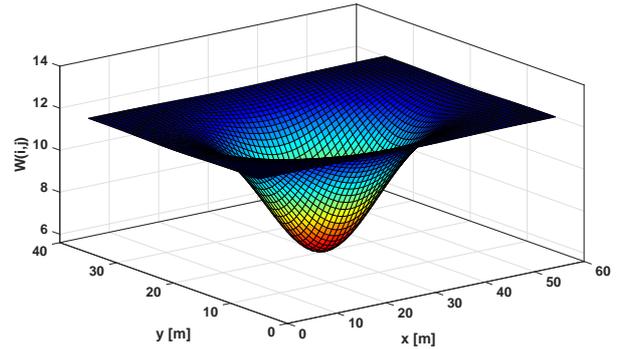}
	\caption{An example of the function $W(i,j)$. The side $S$ of the square cells is 1 m. The size of the room is specified by the values in the x and y axes.}
	\label{fig:surf}
\end{figure}

\begin{enumerate}
	\item \textit{gathering and filtering}:  in this first step, the RSSI values obtained by the beacon are computed. It is necessary to note that in the proposed solution just the values less than the 25th percentile (absolute value) are considered due to multiple reflections and multi-paths. In fact, the lower values of signals could be received due to larger paths and, as a consequence, they are not important because could lead to a wrong estimation of the true distance. Besides, the threshold of the 25th percentile has been chosen as a heuristic solution because it can represent a worthwhile trade-off between the number of values to be filtered and the total number of take-overs. As a result, in a general indoor environment, this pre-filtering step enhances the precision of the received signal.
	\item \textit{estimating of distance}: in the second step, the distance between the unknown node and the beacon ($n$) is estimated ($\hat w_n$).
	\item \textit{development of error map}: in the third step, for each cell, the value $\Arrowvert \hat w_n-w_n(i,j)\Arrowvert$ is associated with it to develop an error map coupled to the anchor $n$ (Figure \ref{fig:errormap}). It is useful to remark that $w_n(i,j)$ is the Euclidean distance between the center of the cell $(i,j)$ and the anchor $n$.
	\item \textit{weighting}: the final step consists in weighting. In fact, the fuzzy reliability index, called $I_n$, is computed to scale the map. $I_n$ is the output value of the FLC 2, as shown in Figure \ref{fig:system-model}, while its input parameters are the reliability index (measured in the previous phase) and a proximity index determined as follows:
	\begin{equation} \label{eq:proxindex}
	\frac{\min\limits_{i} (|RSSI_i|)}{|RSSI_n|}
	\end{equation}
	where the value obtained by the anchor node $n$ is confronted with the biggest value acquired by the unknown node. The input parameters of the FLC 2 are subdivided into 3 fuzzy sets (with triangular membership functions) and are shown in Figure \ref{fig:mf-fuzzy2}, where the membership is realized by normalized values [$0\div 1$]. Furthermore, even in this case, considering the equation \ref{eq:triagular}, the different values of the variables are presented in Table \ref{tab:fuzzy2}. Finally, the output of the FLC 2, reported in Table \ref{tab:output-fuzzy2}, is determined by 9 fuzzy rules and represents the model total reliability index, ranging from $0$ to $1$. For instance, if \textit{Anchor node reliability} is \textit{High} and \textit{Anchor node normalized RSSI} is \textit{Low} then the output value ($I_n$) is $0.3$.
	
	Now, it is possible to outline the the equation of the map, that is the following:
	\begin{equation} \label{eq:map}
	W(i,j)=\displaystyle\sum_{n=1}^{N} I_n\cdot(\hat w_n-w_n(i,j))^2
	\end{equation}
	Regarding the cells, the indexes that decrease at the minimum the error are:
	\begin{equation} \label{eq:index}
	(\tilde{i},\tilde{j})=\arg \min\limits_{i,j} W(i,j)
	\end{equation}
	An example of the function $W(i,j)$ is depicted in Figure \ref{fig:surf}. Finally, the coordinates $(x,y)$ of the position are provided as follows:
	\begin{equation} \label{eq:coordinates}
	\begin{cases}
	x=\tilde{i}\cdot S-S/2\\
	y=\tilde{j}\cdot S-S/2\\
	\end{cases}
	\end{equation}
	where $S$, i.e. the side of the cell, is a project parameter because it is chosen in the implementation phase.
\end{enumerate}

\begin{table}[]
	\centering
	\caption{FLC 2: values of variables used in definition of triangular membership functions.}
	\label{tab:fuzzy2}
	\begin{tabular}{|c|c|c|c|c|}
		\hline
		\textbf{\begin{tabular}[c]{@{}c@{}}Input \\ Variable\end{tabular}}                         & \textbf{Linguistic term} & \textit{\textbf{a}} & \textit{\textbf{m}} & \textit{\textbf{b}} \\ \hline
		\multirow{3}{*}{\begin{tabular}[c]{@{}c@{}}Anchor node \\ reliability\end{tabular}}        & Low                      & 0                   & 0                   & 0.5                 \\ \cline{2-5} 
		& Medium                   & 0                   & 0.5                 & 1                   \\ \cline{2-5} 
		& High                     & 0.5                 & 1                   & 1                   \\ \hline
		\multirow{3}{*}{\begin{tabular}[c]{@{}c@{}}Anchor node \\ normalized \\ RSSI\end{tabular}} & Low                      & 0                   & 0                   & 0.5                 \\ \cline{2-5} 
		& Medium                   & 0                   & 0.5                 & 1                   \\ \cline{2-5} 
		& High                     & 0.5                 & 1                   & 1                   \\ \hline
	\end{tabular}
\end{table}

\begin{table}[]
	\centering
	\caption{FLC 2: inference rules of the model total reliability index.}
	\label{tab:output-fuzzy2}
	\begin{tabular}{|c|c|c|c|c|}
		\hline
		\multicolumn{2}{|c|}{\multirow{2}{*}{\textbf{}}}                                                                  & \multicolumn{3}{c|}{\textbf{Anchor node normalized RSSI}} \\ \cline{3-5} 
		\multicolumn{2}{|c|}{}                                                                          & \textit{Low}     & \textit{Medium}     & \textit{High}    \\ \hline
		\multirow{3}{*}{\textbf{\begin{tabular}[c]{@{}c@{}}Anchor \\ node \\ reliability\end{tabular}}} & \textit{Low}    & 0.001            & 0.3                 & 0.7              \\ \cline{2-5} 
		& \textit{Medium} & 0.01             & 0.4                 & 0.9              \\ \cline{2-5} 
		& \textit{High}   & 0.3              & 0.6                 & 1                \\ \hline
	\end{tabular}
\end{table}

\section{FLC optimization through PSO}
\label{sec:pso} 

Particle Swarm Optimization technique reproduces the actions of crowds of animals to generate the best (or near best) solutions for a function with a specific goal in a constant search domain. PSO consists in a population-based method where a \textit{swarm of particles} goes in the research domain. The positions of this \textit{swarm of particles} designate the applicant solutions of the considered problem. The performance of each particle is strictly related to its position. Its value is determined by a cost function associated with the examined optimization issue. Usually, the opening condition of every single particle is randomly produced. Subsequently, through several iterations, the progress of the particles in the search domain is affected by the present best position, named \textit{personal best position}. Moreover, it depends further on the present best position of the whole particles, denominated \textit{global best position}. As a starting point, it is necessary to consider an individual swarm with a size equals to \textit{K}. Furthermore, this swarm is fully connected in a \textit{N}-dimensional research domain. It is possible to determine and refresh the position and velocity of every single particle as follows:

\begin{equation} \label{eq:velocity}
\begin{split}
\mathbf{v}_{k,n}(t+1) & = w\mathbf{v}_{k,n}(t)+c_1r_1(\mathbf{p}_{k,n}(t)-\mathbf{x}_{k,n}(t))+ c_2r_2(\mathbf{g}_n(t)-\\\mathbf{x}_{k,n}(t))
\end{split}
\end{equation}

\begin{equation}
\label{eq:position}
\mathbf{x}_{k,n}(t+1) = \mathbf{x}_{k,n}(t)+\mathbf{v}_{k,n}(t+1)
\end{equation}

It is necessary to note that $ {1 \le k \le K} $ and $ {1 \le j \le N} $, while $ \textbf{x}_k(t) $ and $ \textbf{v}_k(t) $ are the position and velocity vectors of the \textit{k-th} particle at the \textit{t-th} time step respectively. Moreover, $ \textbf{p}_k(t) $ is the individual best position of the \textit{k-th} particle at the \textit{t-th} time step, $ \textbf{g}(t) $ is the global best position in the entire swarm of particles at the \textit{t-th} time step; $ r_1 $ and $ r_2 $ are causal numbers organized in a uniform way in the range $ \left [ 0,1 \right ] $. Finally, the last parameters to introduce are \textit{w}, i.e. the inertia weight, and $ c_1 $ and $ c_2 $, that are the cognitive coefficients. It is useful to highlight that \textit{w} is employed to achieve a scale of the research domain and performs an essential function in PSO convergence performance. There are various methods of estimating this parameter. However, in most of them, and also in this paper, it can be adjusted to a constant value to decrease the computational load of the algorithm.

The velocity of every single particle, as reported in the eq. \ref{eq:velocity}, is adjusted taking into account the \textit{inertial component}, i.e. its current velocity, the \textit{social component}, and the \textit{cognitive component}. All these parameters are strictly related to the personal best and global best position. The Algorithm 1 represents the pseudo-code of the Particle Swarm Optimization Algorithm introduced in this work. As it is possible to note, in the initial phase the swarm is analyzed and elaborated, initializing the position and velocity of every single particle randomly. Subsequently, the evaluation of the cost function for each particle is carried out. This procedure is performed to achieve the global best position in the swarm. In the next step of the algorithm, the position and velocity of all the particles of the swarm are updated continuously taking into account not only the equations \ref{eq:velocity} and \ref{eq:position} but also the cost function, which is evaluated from time to time. A direct consequence of this mechanism is the upgrade of both the personal and the global best position. In the end, the cycle is terminated if the finish rule is fulfilled. The output of the algorithm, i.e. the solution, consists of the global best position at the last iteration.

\begin{algorithm}
	\caption{Pseudo-code of the PSO}
	\begin{algorithmic} 
		\FOR {(every single particle)}
		\STATE  initialize the velocity and position;
		\STATE  assess the cost function;
		\STATE  estimate the best position (global);
		\ENDFOR
		\REPEAT
		\FOR {every single particle}
		\STATE  refresh the position and velocity based on the equations \ref{eq:velocity} and \ref{eq:position};
		\STATE  assess cost function;
		\STATE  refresh the personal best position and the global best position;
		\ENDFOR
		\UNTIL{finish rule is fulfilled}
		\RETURN the best position (global);
	\end{algorithmic}
\end{algorithm}

\subsection{Particle Swarm Optimization Algorithm}
\label{subsec:opt}

\begin{figure}
	\centering
	\includegraphics[width=3in]{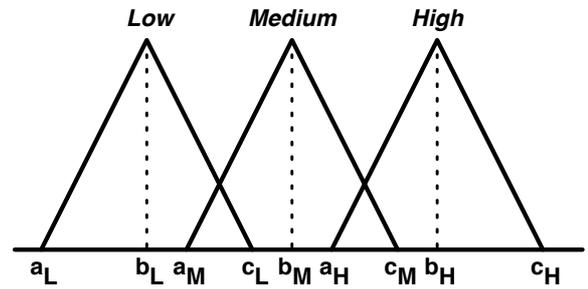}
	\caption{Illustration of generic membership functions (triangular).}
	\label{fig:triangular-pso}
\end{figure}

The architecture of the FLC 1 is depicted in Figure \ref{fig:mf-fuzzy1}. As it is possible to note the inputs of the controller are $Z\%$ and $Kl$, while the output is represented by the \textit{model reliability index}. On the contrary, in the FLC 2, depicted in Figure \ref{fig:mf-fuzzy2}, the controller inputs are the \textit{Anchor node reliability} and the \textit{Anchor node normalized RSSI}, while the output is represented by $I_n$. In both cases, the number of the membership functions is 3 (i.e. \emph{Low}, \emph{Medium}, \emph{High}) for both inputs and outputs. Consequently, as depicted in Tables \ref{tab:output-fuzzy1} and \ref{tab:output-fuzzy2}, the amount of the fuzzy inference rules is 9. As mentioned above, triangular membership functions are taken into account in the approach introduced in this paper, and the goal is to optimize them through the Particle Swarm Optimization. The membership functions can be expressed as in Figure \ref{fig:triangular-pso}. In this paper, the optimization must be simple and should not involve a large computational load. For this reason, it is defined that the parameters $a_L$, $b_M$ and $c_H$, for both inputs and outputs, are fixed. As a consequence, the PSO algorithm has to optimize 18 membership functions parameters. The arrangement of a general particle, for both inputs and outputs, is defined as follows:

\begin{equation}
\label{eq:structure}
\begin{vmatrix} c_L & b_L & a_M & c_M & a_H & b_H\end{vmatrix}
\end{equation}

Examining the Figure \ref{fig:triangular-pso}, it is necessary to specify that in the PSO algorithm introduced in this paper the 6 parameters considered for the optimization of both inputs and outputs need to satisfy not only the following rules but also in the itemized order:

\begin{equation}
\label{eq:constraints}
\begin{matrix} 
1)\;\;a_L<c_L<b_M   &   2)\;\;a_L<b_L<c_L\\
3)\;\;a_L<a_M<c_L   &   4)\;\;b_M<c_M<c_H\\
5)\;\;b_M<a_H<c_M   &   6)\;\;a_H<b_H<c_H 
\end{matrix}
\quad
\end{equation}

In the operation of the PSO algorithm, all the constraints, shown in eq. \ref{eq:constraints}, must be checked in each repetition. Nevertheless, in this paper, the PSO is supported by a proportional method with the aim to decrease the computational cost in such a way to enhance the convergence speed. It is valuable to examine the \textit{n-th} position of the \textit{k-th} particle at the \textit{t-th} iteration to understand the proposed PSO algorithm. The next limitation must be met:

\begin{equation}
\label{eq:domainposition}
\mathbf{x}_{k,n}(t) \in \left [ A_{k,n}(t+1), B_{k,n}(t+1)\right ]
\end{equation}
where the limitations $A_{k,n}(t+1)$ and $B_{k,n}(t+1) $ have previously been refreshed considering the sequence presented in eq. \ref{eq:constraints}. Furthermore, it is helpful to remark that, if necessary, just one of them can be changed.

The key phases of the PSO algorithm presented in this work are the following:

\begin{enumerate}
	\item if the interval $\left [ A_{k,n}(t+1), B_{k,n}(t+1)\right ]$ does not contain the the position $\mathbf{x}_{k,n}(t)$, then the latter is determined proportionally as follows:\\
	if $\mathbf{x}_{k,n}(t)<A_{k,n}(t+1)$, then
	\begin{equation}\begin{split}
	\label{eq:proportion1}
	\mathbf{x}_{k,n}(t)=B_{k,n}(t)+{A_{k,n}(t+1)-B_{k,n}(t) \over B_{k,n}(t)-A_{k,n}(t)}(\mathbf{x}_{k,n}(t)-B_{k,n}(t))
	\end{split}
	\end{equation}
	else if $\mathbf{x}_{k,n}(t)>B_{k,n}(t+1)$, then
	\begin{equation}\begin{split}
	\label{eq:proportion2}
	\mathbf{x}_{k,n}(t)=A_{k,n}(t)+{B_{k,n}(t+1)-A_{k,n}(t) \over B_{k,n}(t)-A_{k,n}(t)}(\mathbf{x}_{k,n}(t)-A_{k,n}(t))
	\end{split}
	\end{equation}
	
	\item the velocity $\mathbf{v}_{k,n}(t) $ is refreshed based on the equation \ref{eq:velocity}. It is useful to highlight that, in the context analyzed in this work, the velocity \textit{n-th} of the \textit{k-th} particle at the \textit{(t+1)-th} iteration is the following:
	\begin{equation}
	\label{eq:domainvelocity}
	\mathbf{v}_{k,n}(t+1) \in \left [\mathbf{v}_{k,n}^{(min)}(t+1), \mathbf{v}_{k,n}^{(max)}(t+1)\right ]
	\end{equation}
	where $\mathbf{v}_{k,n}^{(min)}(t+1)$  and $\mathbf{v}_{k,n}^{(max)}(t+1)$  are determined ad follows:
	
	\begin{equation}\begin{split}
	\label{eq:minvelocity}
	\mathbf{v}_{k,n}^{(min)}(t+1) = w\mathbf{v}_{k,n}(t)+c_1r_1(\mathbf{p}_{k,n}(t)-B_{k,n}(t+1))\\+c_2r_2(\mathbf{g}_n(t)-B_{k,n}(t+1))
	\end{split}
	\end{equation}
	
	\begin{equation}\begin{split}
	\label{eq:maxvelocity}
	\mathbf{v}_{k,n}^{(max)}(t+1) = w\mathbf{v}_{k,n}(t)+c_1r_1(\mathbf{p}_{k,n}(t)-A_{k,n}(t+1))+\\c_2r_2(\mathbf{g}_n(t)-A_{k,n}(t+1))
	\end{split}
	\end{equation}	
	
	\item the position $\mathbf{x}_{k,n}(t)$ is refreshed based on the equation \ref{eq:position}. If the interval interval $\left [ A_{k,n}(t+1), B_{k,n}(t+1)\right]$ does not contain the position $\mathbf{x}_{k,n}(t+1) $, at first, the minimum and the maximum values of the velocity are estimated based on the equation \ref{eq:minvelocity} and \ref{eq:maxvelocity}, and, subsequently, the position $\mathbf{x}_{k,n}(t+1)$ is estimated proportionally as follows:
	
	if $\mathbf{x}_{k,n}(t+1)<A_{k,n}(t+1)$, then
	\begin{equation}
	\label{eq:proportion3}
	\mathbf{x}_{k,n}(t+1)=\mathbf{x}_{k,n}(t+1)+{\mathbf{v}_{k,n}(t+1) \over \mathbf{v}_{k,n}^{(min)}(t+1)}(A_{k,n}(t)-\mathbf{x}_{k,n}(t))
	\end{equation}
	
	else if $\mathbf{x}_{k,n}(t)>B_{k,n}(t+1)$, then
	\begin{equation}
	\label{eq:proportion4}
	\mathbf{x}_{k,n}(t+1)=\mathbf{x}_{k,n}(t+1)+{\mathbf{v}_{k,n}(t+1) \over \mathbf{v}_{k,n}^{(max)}(t+1)}(B_{k,n}(t)-\mathbf{x}_{k,n}(t))
	\end{equation}
	
\end{enumerate}

\subsection{Particle Swarm Optimization Performance}
\label{subsec:psoperf}

\begin{figure}
	\centering
	\includegraphics[width=3.5in]{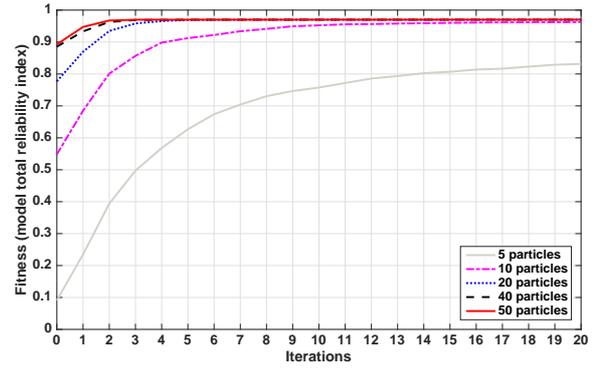}
	\caption{Performance of the proposed Particle Swarm Optimization.}
	\label{fig:psoperf}
\end{figure}

The performance of the suggested Particle Swarm Optimization method is addressed in this section. For the sake of simplicity, only the performance of the PSO related to the FLC 2 is shown. The inputs of the FLC 2 (Figure \ref{fig:system-model}) are the \textit{Anchor node reliability} and the \textit{Anchor node normalized RSSI}, while the output is the model total reliability index. This value has been considered as the fitness function. As a consequence, the Particle Swarm Optimization algorithm has been implemented to obtain its maximum value. The Particle Swarm Optimization performance has been analyzed taking into account swarms with different sizes, i.e. $k={5,10,20,40.50}$. Figure \ref{fig:psoperf} presents the model total reliability index reached by Particle Swarm Optimization method presented in this paper. The values, obtained through simulations carried out with Matlab, have been averaged over $250$ executions for $20$ iterations. It is beneficial to perceive that, in each simulation, the algorithm has been initialized randomly, the cognitive coefficients have been established to $c1 = c2 = 1.47$, the value of the inertia weight has been $w = 0.74$, and the state of $20$ iterations has been considered as the the end check. The achieved results show that the performance of the proposed algorithm is enhanced by enlarging the swarm concerning size. In fact, the simulation by using $k=20$, $k=40$ and $k=50$ achieves the convergence, i.e. the maximum value, after about $3$ iterations. However, it is useful to perceive that the PSO algorithm developed with $k=50$ holds a greater speed, regarding the achievement of the convergence (i.e. $2$ iterations), compared to smaller sized swarms. In fact, $k=50$ examines in the best way the research domain by treating further particles in every single iteration.

\section{Performance Evaluation}
\label{sec:performance}

A testbed scenario, composed of different LED lamps and an optical receiver, was developed to validate the proposed fuzzy-based solution optimized through the PSO algorithm. The lamps used in the testbed had 18 white LEDs to provide illumination of 60 lux. The LED (NBL-R3W) has a viewing angle of 30 degrees, and the standard light power is 5.0 cd. The LED lamps used in the testbed scenario were managed by an 8-bit microcontroller (ATmega128). The receiver used a low-cost photodiode (SFH-213) to estimate the intensity of the light source. The radiant sensitive area of the photodiode is $1 mm^2$ while the half angle is $\pm10$ degree. The measures were obtained by persisting $30$ seconds in each position, both in offline and online stages (Figure \ref{fig:system-model}), in an environment whose area is $100 m^2$. In other words, between $4$ and $10$ anchor nodes (LED lamps) were used. In each experimental scenario, $20$ different and known positions were chosen. Moreover, the performance of the proposed fuzzy-based solution was then compared with those of MinMax, Maximum Likelihood and Trilateration.

\begin{figure}
	\centering
	\subfigure[]
	{\includegraphics[width=1.5in]{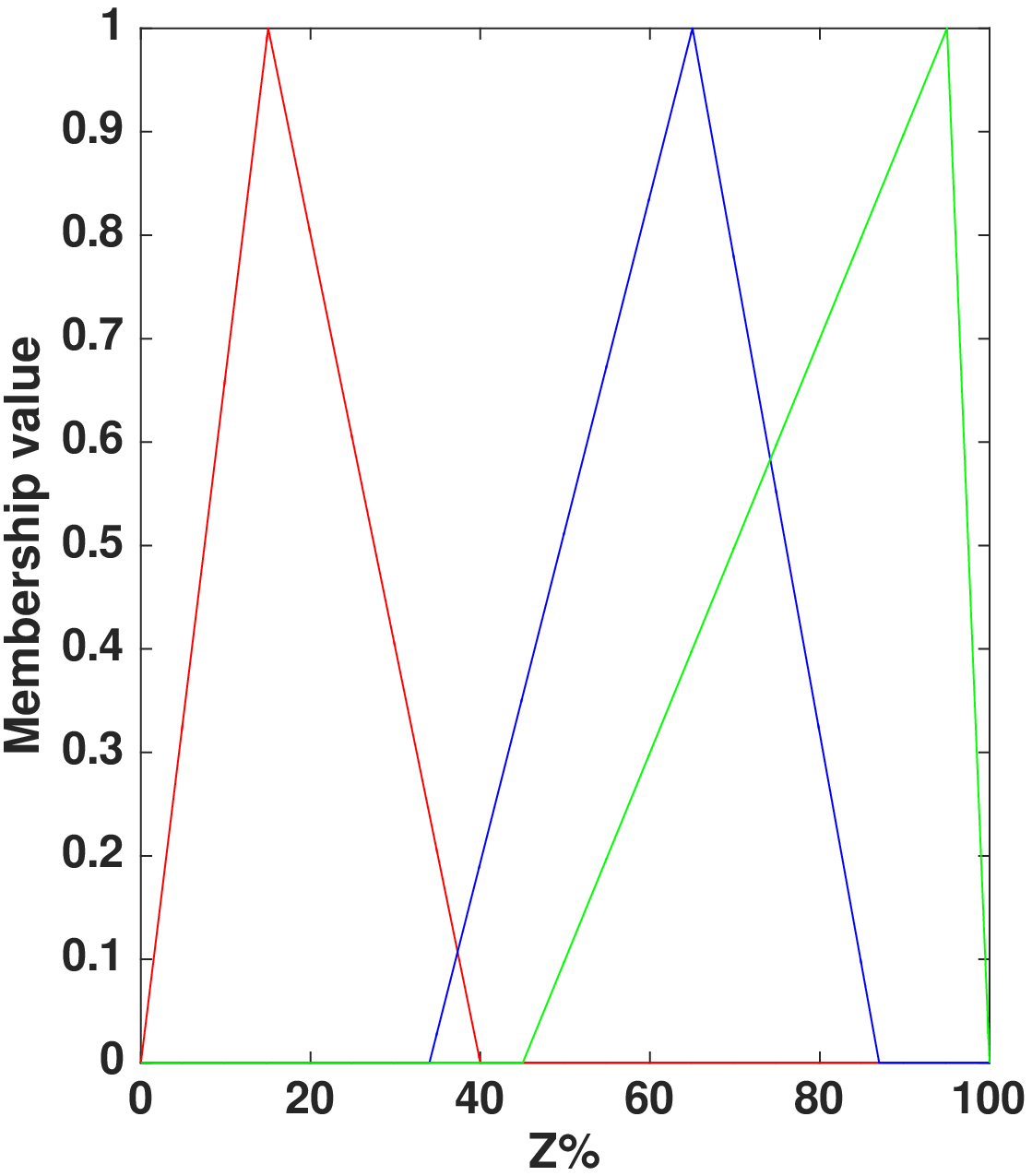}}
	\subfigure[]
	{\includegraphics[width=1.5in]{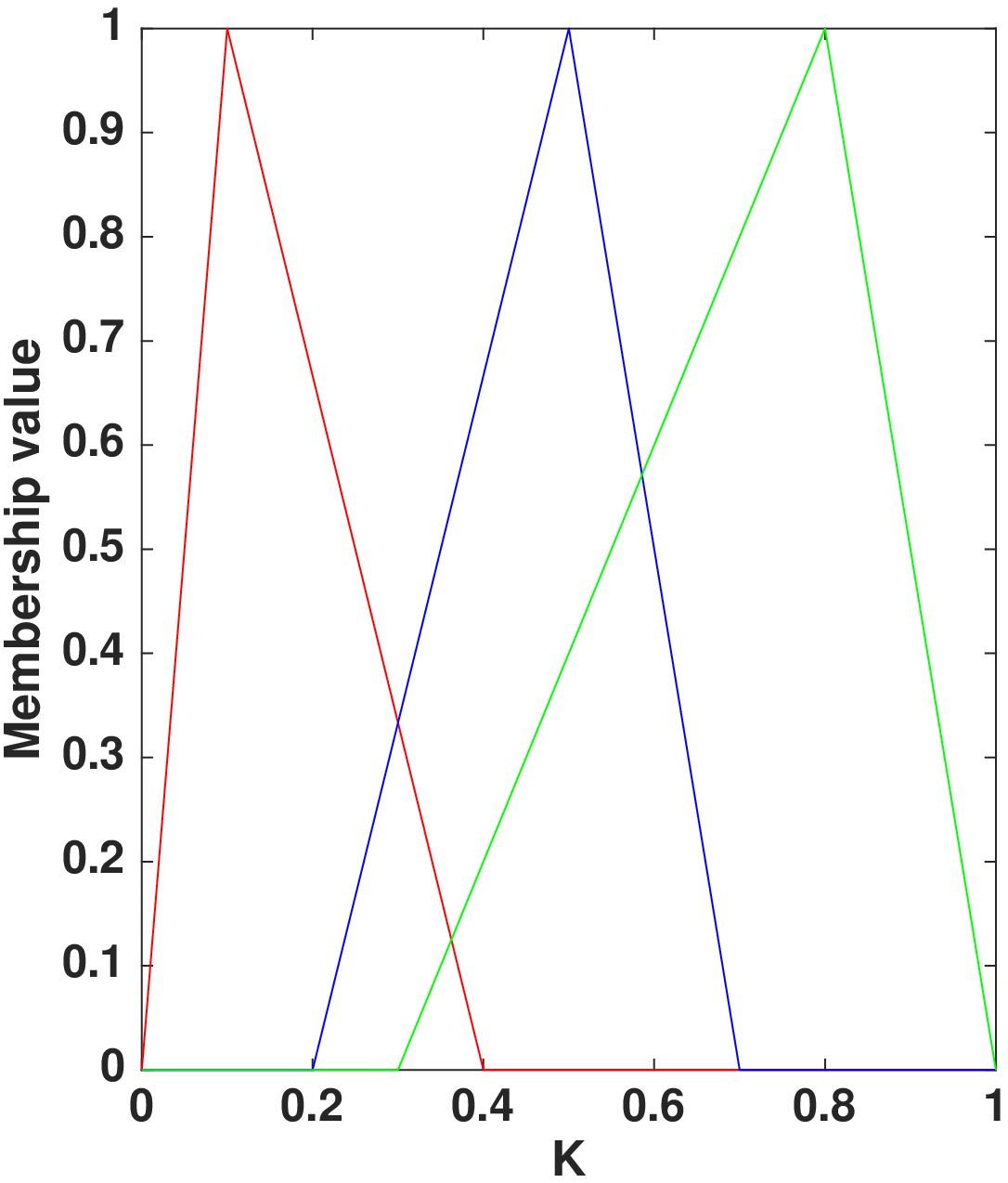}}
	\subfigure[]
	{\includegraphics[width=1.5in]{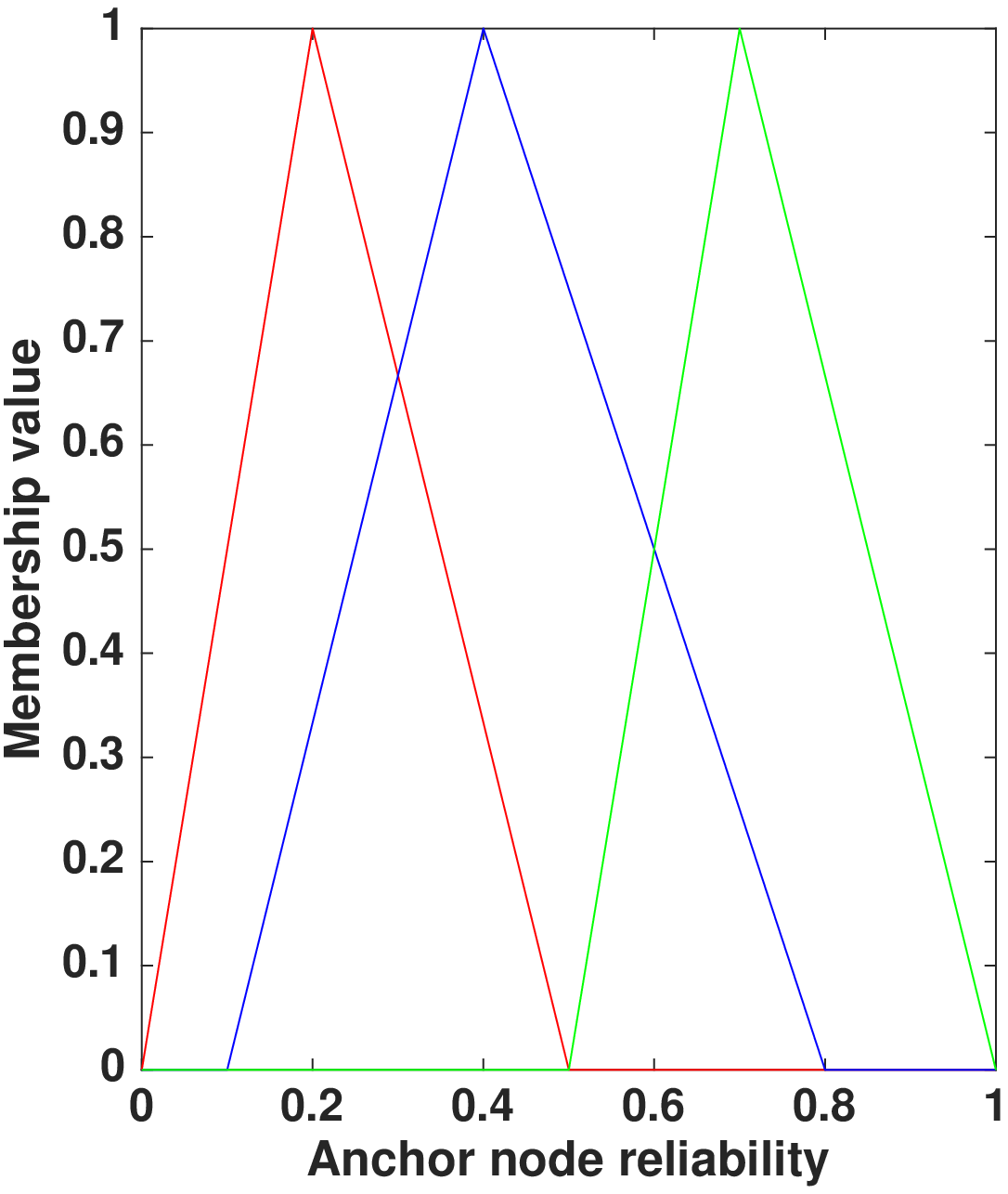}}
	\subfigure[]
	{\includegraphics[width=1.5in]{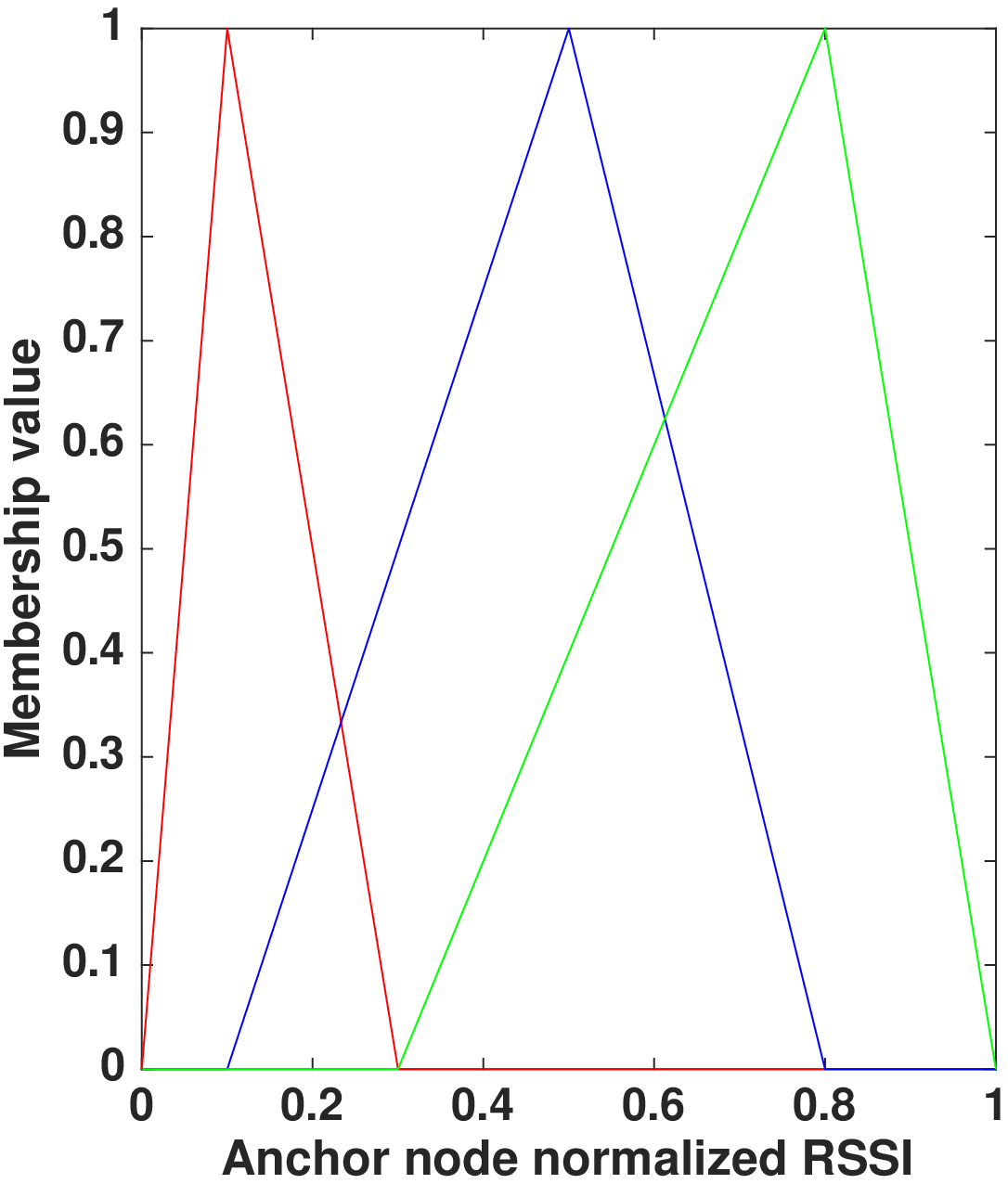}}
	\caption{Membership functions achieved by the PSO algorithm ($50$ particles and $20$ iterations).}
	\label{fig:mf-fuzzy-pso}
\end{figure}

Several measurements were conducted both to validate the method presented in this paper, i.e. indoor localization through VLC, and, principally, to examine the various solutions of the Particle Swarm Optimization by diversifying the swarm size regarding the number of the particles. As mentioned before, the goal of the PSO method is to optimize both FLCs regarding their membership functions, modifying their range. It is essential to examine the membership functions presented in Figures \ref{fig:mf-fuzzy1} and \ref{fig:mf-fuzzy2} to determine the run time of the PSO. These membership functions were statically assigned for both FLCs (Section \ref{sec:system-model}). Using PSO, the results achieved using $50$ particles and $20$ iterations are shown in Figure \ref{fig:mf-fuzzy-pso}. In this case, the range of the membership functions was considerably varied compared to those depicted in Figures \ref{fig:mf-fuzzy1} and \ref{fig:mf-fuzzy2}. Nevertheless, in almost all cases, PSO offers better performance. For simplicity, we presented the range of the membership functions achieved with $50$ particles and $20$ iterations since these values yielded the best performance. 

\subsection{Findings}
\label{subsec:results}

Statistical metrics and the the Cumulative Distribution Function (CDF) \citep{rssi-refref,cdf} of localization error were taken into account when evaluating the performance. The CDF ($F(e)$) of a localization error $e$, where $f(e)$ denotes a probability density function, is defined as follows:

\begin{equation}
\label{eq:cdf}
F(e)=\int_0^e f(x)\,\mathrm{d}x \quad (x\geq 0)
\end{equation}
In fact, considering the CDF of localization error, it is feasible to determine the localization error at an assigned confidence level (for instance $10\%$, $50\%$, $90\%$). The performance of three popular methods (i.e. MinMax, Maximum Likelihood, and Trilateration) were also evaluated. We remark the performance on load and computational complexity is not included in this paper since the setting does not influence the algorithmic complexity \citep{computing}.

\begin{table}[]
	\centering
	\caption{Comparison among localization approaches (error in meters).}
	\label{tab:locerror}
	\begin{tabular}{|c|c|c|c|}
		\hline
		\textit{\textbf{Algorithm}} & \textbf{AE} & \textbf{ME} & \textbf{SD} \\ \hline
		MinMax                      & 1.94        & 1.91        & 1.21        \\ \hline
		Maximum Likelihood          & 2.03        & 1.97        & 1.24        \\ \hline
		Trilateration               & 2.28        & 2.05        & 1.31        \\ \hline
		Fuzzy without PSO           & 1.88        & 1.78        & 0.99        \\ \hline
		Fuzzy-PSO: 5 particles      & 1.97        & 2.02        & 1.23        \\ \hline
		Fuzzy-PSO: 10 particles     & 1.71        & 1.65        & 1.01        \\ \hline
		Fuzzy-PSO: 20 particles     & 1.44        & 1.33        & 0.85        \\ \hline
		Fuzzy-PSO: 40 particles     & 0.95        & 0.67        & 0.57        \\ \hline
		Fuzzy-PSO: 50 particles     & 0.75        & 0.43        & 0.35        \\ \hline
	\end{tabular}
\end{table}

For every record of the obtained data at a target position, the algorithms were applied to determine the position and to compare this value with its real value for error evaluation. The achieved performance of the localization approaches is shown in Table \ref{tab:locerror}. The proposed fuzzy solution (with and without the PSO) was able to achieve better performance, in comparison to the other three algorithms, in terms of Average Error (AE), Median Error (ME) and Standard Deviation (SD). In detail, the excellent outcomes are obtained by employing the Particle Swarm Optimization considering a swarm with more than $5$ particles. This is because of the use of the triangular membership functions determined by the PSO with $50$ particles. We also determined that any more than $50$ particles will offer only modest improvement, particularly if more than $50$ particles are used. 

\begin{figure}
	\centering
	\includegraphics[width=3.5in]{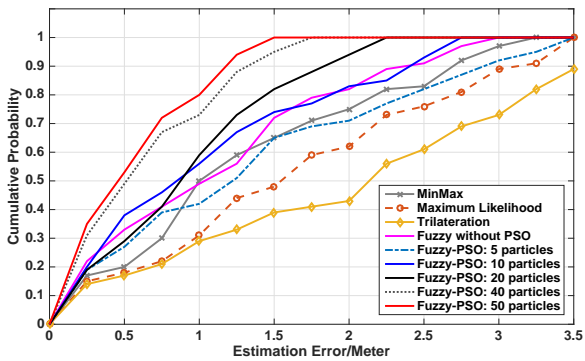}
	\caption{Cumulative probability function.}
	\label{fig:cdfperf}
\end{figure}

\begin{figure}
	\centering
	\includegraphics[width=3.5in]{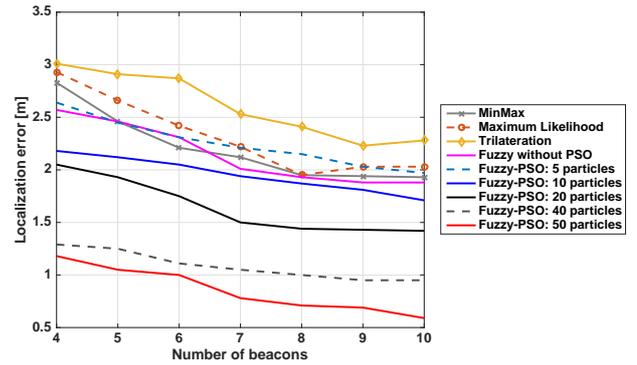}
	\caption{Errors depending on beacon number.}
	\label{fig:beaconsperf}
\end{figure}

Figure \ref{fig:cdfperf} shows the cumulative probability function of the estimation error in the four algorithms. Analyzing the results depicted in Figure \ref{fig:cdfperf}, it is clear that the proposed fuzzy solution outperformed MinMax, Maximum Likelihood, and Trilateration algorithms. For instance, the estimation error of the fuzzy-PSO with $50$ particles was less than those of MinMax, Maximum Likelihood, and Trilateration, at both $50\%$ and $90\%$ confidence levels. In fact, the estimation of Error/Meter could be at most equal to $1.5$ meters. This value (i.e. the maximum for fuzzy-PSO with $50$ particles) is satisfactory in a $100m^2$ environment. On the contrary, with $5$ particles in the PSO, the worst performance, compared to the use of a greater number of particles, were obtained. However, in all other cases, even without PSO, the performance were always better than those achieved with MinMax, Maximum Likelihood, and Trilateration. It is necessary to highlight that the average error fluctuated based on the number of beacons, as depicted in Figure \ref{fig:beaconsperf}. In fact, there is a clear error reduction when the number of anchors increased from $6$ to $7$, and when there is a large beacon density, the improvement is almost negligible. Our findings echoed those in the literature (i.e. an improvement in precision with a major density) \citep{accuracy1,accuracy2}.

\section{Conclusions} \label{sec:conclusions} 
In this paper, a novel solution for the indoor localization, based on the combined use of Fuzzy Logic and Visible Light Communications, was introduced. We demonstrated via a real testbed scenario that the proposed system can achieve optimal FLCs parameters due to the optimization of membership functions. In fact, their range can be adjusted to produce optimal localization reliability. In our approach, we also applied PSO technique. 

Future research includes expanding the scope of the evaluation, such as the number of algorithms to be compared against and a broader set of environmental configurations. 
\\ \\
%
\textbf{Compliance with ethical standards} \\ \\
\textbf{Conflict of interest} All authors declare that they have no conflict of interest.\\ \\
\textbf{Ethical standard} This article does not contain any studies with human participants or animals performed by any of the authors.

\bibliographystyle{spbasic}      
\bibliography{IEEEfull}   


\end{document}